\documentclass[journal]{IEEEtran}
\IEEEoverridecommandlockouts
\usepackage{cite}
\usepackage{amsmath,amssymb,amsfonts}
\usepackage{algorithmic}
\usepackage{graphicx}
\usepackage{textcomp}
\usepackage{acronym}
\usepackage{url}
\usepackage{tabularx}
\usepackage{booktabs}
\usepackage{multirow}
\usepackage{siunitx}
\usepackage{wasysym}

\usepackage{prettyref}
\newrefformat{chap}{Chapter~\ref{#1}}
\newrefformat{sec}{Sec.~\ref{#1}}
\newrefformat{subsec}{Sec.~\ref{#1}}
\newrefformat{eq}{Eq.~(\ref{#1})}
\newrefformat{fig}{Fig.~\ref{#1}}
\newrefformat{ex}{Example~\ref{#1}}
\newrefformat{list}{List~\ref{#1}}
\newrefformat{tab}{Table~\ref{#1}}
\newrefformat{enum}{Case~(\ref{#1}.)}
\newrefformat{def}{Definition~\ref{#1}}
\newcommand{\pref}[1]{\prettyref{#1}}

\newcommand{\eg}{e.\,g., }

\newcommand{\ie}{i.\,e., }

\newcommand{\cf}{cf.~}

\acrodef{dlt}[DLT]{Distributed Ledger Technology}
\acrodef{dos}[DoS]{Denial-of-Service}
\acrodef{ddos}[DDoS]{Distributed Denial-of-Service}
\acrodef{ics}[ICS]{Industrial Control System}
\acrodef{ids}[IDS]{Intrusion Detection System}
\acrodef{iot}[IoT]{Internet of Things}
\acrodef{iiot}[IIoT]{Industrial Internet of Things}
\acrodef{it}[IT]{Information Technology}
\acrodef{jtag}[JTAG]{Joint Test Action Group}
\acrodef{m2m}[M2M]{Machine-to-Machine}
\acrodef{mitm}[MitM]{Man-in-the-Middle}
\acrodef{mud}[MUD]{Manufacturer Usage Description}
\acrodef{nist}[NIST]{National Institute of Standards and Technology}
\acrodef{ot}[OT]{Operational Technology}
\acrodef{owasp}[OWASP]{Open Web Application Security Project}
\acrodef{plc}[PLC]{Programmable Logic Controller}
\acrodef{sdn}[SDN]{Software-Defined Networking}

\begin{document}

\title{Challenges and Opportunities in Securing the Industrial Internet of Things}

 \author{Martin~Serror,
         Sacha~Hack,
         Martin~Henze,
         Marko~Schuba,
         and~Klaus~Wehrle,~\IEEEmembership{Member,~IEEE}%
\thanks{Manuscript received July 2, 2019; revised December 18, 2019 and May 11, 2020; accepted September 5, 2020. This research was supported by the research training group ``Human Centered System Security'' sponsored by the German federal state of North Rhine-Westphalia.}
\thanks{M.\ Serror and K.\ Wehrle are with the Chair of Communication and Distributed Systems at RWTH Aachen University, Germany (e-mail: \{serror,wehrle\}@comsys.rwth-aachen.de).}%
\thanks{S.\ Hack and M.\ Schuba are with FH Aachen University of Applied Sciences, Germany (e-mail: \{s.hack,schuba\}@fh-aachen.de).}%
\thanks{M.\ Henze is with the Fraunhofer Institute for Communication, Information Processing and Ergonomics FKIE, Germany (e-mail: martin.henze@fkie.fraunhofer.de).}
}

\makeatletter
\def\ps@IEEEtitlepagestyle{%
\def\@oddfoot{\parbox{\textwidth}{\footnotesize
This is the author's version of an article that has been published in IEEE Transactions on Industrial Informatics (Volume: 17, Issue: 5, May 2021).
Changes were made to this version by the publisher prior to publication.
The final version of record is available at http://dx.doi.org/10.1109/TII.2020.3023507.\\
Copyright (c) 2020 IEEE. Personal use is permitted. For any other purposes, permission must be obtained from the IEEE by emailing pubs-permissions@ieee.org.\vspace{0.2em}}
}%
}
\makeatother

\maketitle

\begin{abstract}
Given the tremendous success of the Internet of Things in interconnecting consumer devices, we observe a natural trend to likewise interconnect devices in industrial settings, referred to as Industrial Internet of Things or Industry~4.0.
While this coupling of industrial components provides many benefits, it also introduces serious security challenges.
Although sharing many similarities with the consumer Internet of Things, securing the Industrial Internet of Things introduces its own challenges but also opportunities, mainly resulting from a longer lifetime of components and a larger scale of networks.
In this paper, we identify the unique security goals and challenges of the Industrial Internet of Things, which, unlike consumer deployments, mainly follow from safety and productivity requirements.
To address these security goals and challenges, we provide a comprehensive survey of research efforts to secure the Industrial Internet of Things, discuss their applicability, and analyze their security benefits.
\end{abstract}


\section{Introduction}
\label{sec:intro}

\IEEEPARstart{T}{he} proliferation of the \ac{iot}~\cite{AIMo10} has led to a vast amount of interconnected devices.
These devices, typically ranging from small sensors to complex controllers and home appliances, offer all sorts of monitoring and control services to enhance and automate daily tasks.
Due to decreasing costs for hardware and software, more and more of these Internet-enabled devices end up in private homes, contributing to a heterogeneous landscape of distributed computing devices.
While initially focused on consumers, the success of the \ac{iot} has recently spread to other domains, such as the industrial sector, where a similar trend towards connecting previously isolated components with each other and with the Internet is emerging.
This trend is commonly referred to as \emph{Industry~4.0}, \ie the fourth industrial revolution, or as \emph{\ac{iiot}}~\cite{XHLi14, WSJa17}. 
The advantages of coupling industrial components are convincing: Increased flexibility and enhanced process optimization lead to reduced deployment and maintenance costs while providing new services and customize processes to manufacturers, operators, and customers.

On the downside, however, the \ac{iot} bears severe security and privacy risks for its users, an issue that has extensively been surveyed by the research community~\cite{MYAZ15, YWY+17, CLB+17, ALAM19}.
Major reasons for the inherent security flaws in the \ac{iot} are missing or poorly implemented security features, which often remain unpatched due to complicated or lack of updates~\cite{YSS+15}.
Besides, typical users are not aware of the security risks introduced by their \ac{iot} devices and do, moreover, not know how to configure their networks securely~\cite{LiNe17}.
As a result, \ac{iot} devices are increasingly victims of attacks, \eg Bashlite~\cite{MAF+18} and Mirai~\cite{AAB+17}.

Initial research results regarding the security of Industry~4.0 indicate that \ac{iiot} devices are equally affected by vulnerabilities \cite{WHA+16,SWWa15,SSH+18}, drawing a similarly bleak picture for the security in current \ac{iiot} deployments.
Moreover, successful attacks on the availability or operational safety of industrial facilities are typically devastating.
Prominent examples include the attack on a German steel mill in 2014~\cite{LACo14} or the Ukraine power grid in 2015~\cite{WOGS17}.
Indeed, depending on the targeted facility, outages may affect not only a single company, but also clients and suppliers, or even a country's critical infrastructure.

Importantly, the challenges for realizing security in industrial environments significantly differ from the challenges faced in consumer settings.
A significant difference between the two domains is, e.g., the longer lifetime of industrial devices compared to consumer devices requiring the subsequent provision of security measures and prolonged patch management.
Likewise, \ac{iiot} networks are typically larger in scale compared to consumer \ac{iot} deployments consisting only of few devices.
Moreover, the increased and dynamic interconnection between \ac{iiot} devices makes the implementation of a secure network architecture based on segregation more difficult, especially when the offered services evolve.

To establish security in the \ac{iiot}, it makes sense first to acquire a profound understanding of the security deficits and weaknesses in the current consumer \ac{iot}.
This allows us to rectify well-known security flaws even before similar devices are deployed in Industry~4.0 scenarios.
However, not all existing security measures for the consumer \ac{iot} carry over to the industrial domain.
This is mainly due to the differences in use and deployment, as well as in the different security and privacy threats that the consumer and the industrial world are facing.
Consequently, we require a dedicated survey of the distinct security challenges and goals in the \ac{iiot} and Industry~4.0 as a foundation to identify security approaches specifically tailored to the characteristics of the \ac{iiot}.

In this paper, we hence assess the security challenges of the emerging \ac{iiot} regarding vulnerabilities, risks, and threats.
Compared to other \ac{iiot} security surveys (\cf \pref{sec:related-work}), we provide a mapping between countermeasures and identified challenges, thereby specifically considering the unique characteristics of \ac{iiot} deployments, such as long-lived components and increased connectivity.
Our work thus complements existing related work by merging previously independently considered approaches into a comprehensive \ac{iiot} security survey.
To this end, we provide the following original contributions:
\begin{enumerate}
\item We begin with a security analysis based on the well-researched consumer \ac{iot}, which allows us to identify the main differences regarding security when moving towards industrial deployments in the \ac{iiot}.
\item We use this analysis as a foundation to derive and discuss the distinct security goals and challenges faced by the \ac{iiot}, which are heavily influenced by safety and productivity requirements.
\item Based on our results, we survey current research and best practices for improving security in \ac{iiot}.
For each presented approach, we discuss its applicability in \ac{iiot} scenarios and evaluate its security benefits.
\end{enumerate}
The remainder of this paper is structured as follows:
In \pref{sec:iot-sec}, we summarize the security issues and countermeasures for the consumer \ac{iot}. 
Then, in \pref{sec:towards-iiot}, we point out the differences between consumer and industrial \ac{iot}, allowing us to identify the unique \ac{iiot} security goals and challenges in \pref{sec:iiot-risks}.
Afterward, we provide a detailed survey on securing the \ac{iiot} in \pref{sec:iiot-sec}.
Subsequently, we outline a discussion of related work in \pref{sec:related-work}.
We conclude this paper with a summary and an outlook on further research directions in \pref{sec:conclusion}.

\section{IoT Security Threats}
\label{sec:iot-sec}

The security and privacy issues of the (consumer) \ac{iot} are well-researched, out of which many publications have emerged recently.
A common approach is to first classify attacks according to the typical architectural layers of the \ac{iot} in order to provide suited countermeasures at each layer.
In the following, we first summarize related surveys on \ac{iot} security regarding attacks (\cf \pref{sec:iot-sec-attacks}) and countermeasures (\cf \pref{sec:iot-sec-countermeasures}) before discussing their implications for \ac{iiot} security (\cf \pref{sec:iot-sec-discussion}).

\subsection{Attacks}
\label{sec:iot-sec-attacks}

\begin{figure}
	\centering
	\includegraphics[width=\columnwidth]{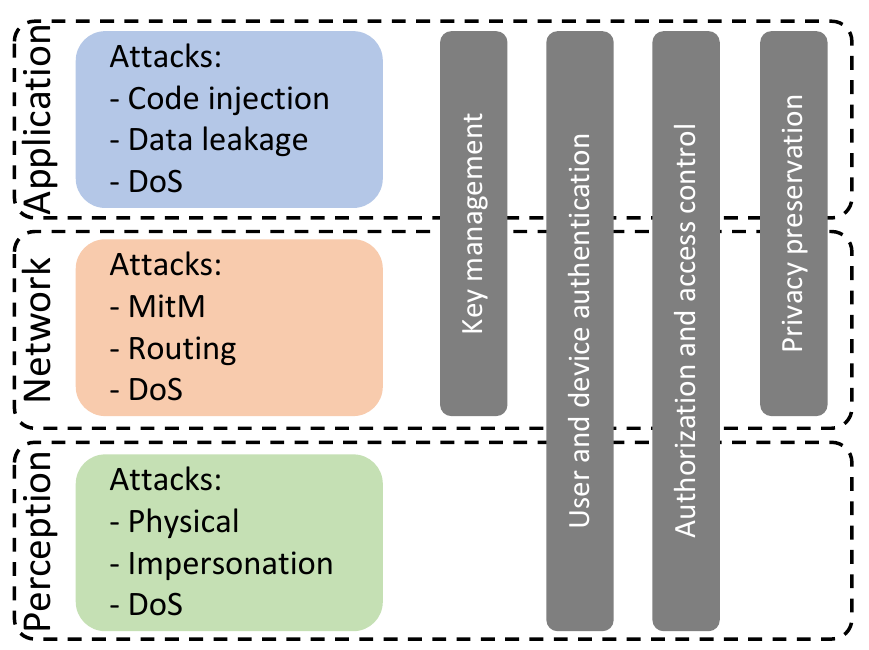}
	\caption{IoT reference layers with attack categories and countermeasures.}
	\label{fig:iot}
\end{figure}

As a general taxonomy for \ac{iot} attacks, a layered model has prevailed, which, in most cases, consists of three \ac{iot} layers, namely \emph{perception}, \emph{network}, and \emph{application}, \eg~\cite{MYAZ15, KVSr16, ELBe16, YWY+17, DeVi17, FPAF18}.
\pref{fig:iot} depicts the different layers as well as common attacks that occur at the respective layer, where some attacks may also occur on multiple layers.
In the following, we shortly explain the scope of each layer and the most common types of \ac{iot} attacks, respectively.

\subsubsection{Perception Layer}
The perception layer describes the interface between the physical world and \ac{iot} devices.
It comprises all sensing activities generating data, as well as actuators interacting with their environment.
This layer is further characterized by sending collected data and receiving commands through the network layer.
Attacks on this layer can be classified as follows.

\textbf{Physical attacks.}
This attack category includes all attacks that are directed to \ac{iot} hardware components.
A typical assumption is that the attacker has physical access to the device and either replaces or damages components to gain access to sensitive information such as user passwords or to disable certain functions, \eg  by accessing the device through its \acs{jtag} interface~\cite{ViLe18}.

\textbf{Impersonation attacks.}
Prominent examples are \emph{spoofing}, \ie assuming a false identity, or \emph{Sybil} attacks, \ie creating a large number of fake identities. 
Such attacks are successfully carried out when authentication mechanisms are missing and particularly successful in the setup phase of \ac{iot} devices, \eg when an attacker advertises as a false WiFi access point~\cite{ZWYZ17}.

\subsubsection{Network Layer}

The purpose of the network layer is to interconnect \ac{iot} devices with each other and with the Internet, \ie transporting collected data from the perception layer to the application layer and vice versa.
Therefore, the key technologies are wireless communication protocols, as well as gateways connecting local \ac{iot} networks to larger core networks such as the Internet.
Attacks on this layer may, therefore, also extend to network components and middleboxes.

\textbf{\acf{mitm}.}
\ac{mitm} attacks aim to eavesdrop on sensitive information by intercepting communication between two legitimate partners and sometimes even to manipulate the data before forwarding it to the other party. 
Such attacks often also include \emph{replay} attacks, \ie repeating or delaying eavesdropped messages.

\textbf{Routing attacks.}
Here the attacker threatens the availability of the local communication network either by propagating false routing information or by interrupting communication flows.
Prominent examples are, \eg \emph{sinkhole} attacks, where an attacker drops messages in a multi-hop network, and \emph{selective forward} attacks, where an attacker only drops some (selected) messages.

\subsubsection{Application Layer}

The application layer provides the different services that constitute the \ac{iot} applications.
It relies on the data of the perception layer and the communication enabled by the network layer.
The application layer may evolve, based on software, with little or no changes to the other layers.
Securing this layer is thus especially challenging, since software changes may introduce new vulnerabilities at any time.
In particular, frequent attacks on this layer can be summarized as follows.

\textbf{Malicious code injection.}
When an attacker successfully injects malicious code exploiting a vulnerability, the attacker gains control over the infected device.
Possible motivations to misuse the existing computing and communication resources are: (i) Recruiting devices for a \emph{\ac{ddos}} attack, \ie an attacker controls a large number of \ac{iot} devices to distributively attack a server by overloading it with requests. (ii) Performing a \emph{cryptojacking} attack, \ie unauthorized use of someone else's computational resources to mine a cryptocurrency. (iii) To extort money with \emph{ransomware} that encrypts data storage of a victim to press money.

\textbf{Data leakage.}
Attackers may use software and service vulnerabilities to steal personal information, \eg by attacking poorly configured cloud services.

\subsubsection{Multi-layer Attacks}

Furthermore, some attacks may occur on multiple layers of the \ac{iot} reference design with varying impacts.
In the following, we present denial-of-service attacks as a prominent example of this category. 

\textbf{\acf{dos}.}
From the user's perspective, a \ac{dos} attack refers to making a resource or service unavailable, typically by overloading the involved machines with requests.
From an architectural perspective, however, it makes sense to differentiate on which \ac{iot} reference layer a \ac{dos} attack occurs to take appropriate countermeasures.
In particular, when centralized entities at the respective layers operate as a single point of failure, they are an especially vulnerable target of \ac{dos} attacks and thus require extra protection.

On the perception layer, \ac{dos} attacks mostly refer to jamming attacks interrupting wireless communication or sensing abilities of \ac{iot} devices.
Furthermore, such attacks target the availability of single \ac{iot} devices, \eg by overloading the resource-constraint processors of \ac{iot} devices.
On the network layer, in turn, \ac{dos} attacks target the communication infrastructure resulting in temporarily disconnected devices, \eg by intentionally overloading routers.
Finally, on the application layer, \ac{dos} attacks refer to attacking critical services, \eg by flooding them with requests, to compromise the availability of (multiple) industrial processes relying on this service.

Security threats for the (consumer) \ac{iot} are thus present at each layer of the \ac{iot} reference model.
Existing research addresses these issues by proposing different security measures, which we deepen in the following.

\subsection{Countermeasures}
\label{sec:iot-sec-countermeasures}

To mitigate the presented attacks, several countermeasures need to be considered for which different taxonomies have been proposed, \eg by \cite{NLOu15, FPAl17, DZHe18, ARCr18}.
Based on these results, we use a comprehensive taxonomy of \ac{iot} security protocols classifying the approaches according to \emph{key management}, \emph{user/device authentication}, \emph{access control}, and \emph{privacy preservation}, which we shortly present in the following.
Moreover, we depict their relation to the \ac{iot} reference layers and the attacks in \pref{fig:iot}.

\textbf{Key management} is a prerequisite for encrypted communication and authentication, where the challenge lies in the scalability and heterogeneity of \ac{iot} deployments. 
In general, public-key mechanisms are easier to manage than symmetric schemes but require more computational resources~\cite{NLOu15}.

\textbf{User and device authentication} allows users and devices to prove that they are who they claim to be. 
For users, typically, multiple factors can be used for authentication.
For devices, in turn, contextual information may be used, such as fingerprinting~\cite{DZHe18}.

\textbf{Authorization and access control} restricts access of users and devices to the required resources and services. 
This can be achieved, for example, with rules and policies, which, however, rely on authentication.

\textbf{Privacy preservation} refers to protecting sensitive personal information, such as medical data, from others.
Besides encryption of data and communication, this also comprises privacy-preserving processing of data, such as secure multi-party computations~\cite{Oles09}.

\subsection{Discussion}
\label{sec:iot-sec-discussion}

The presented results show that \ac{iot} security is a well-researched topic regarding possible risks and countermeasures. 
Since attacks on the (consumer) \ac{iot} are, in general, equally threatening the \ac{iiot}, both domains share common security challenges.
These challenges, however, have not yet been solved comprehensively and are still subject to active research~\cite{HCS+19,NBC+19,KhSa18}. 
Nevertheless, for \ac{iiot} security, which is currently in its infancy, researchers and developers can still rely on a large corpus of tools and methodologies that exist for the consumer \ac{iot}.
Even more importantly, they can learn from the experiences and mistakes that were made in the consumer \ac{iot} and build on the well-researched results of \ac{iot} security.

Beyond this common approach and as we show in the further course of this paper, there are fundamental differences between the consumer and the industrial \ac{iot}, which significantly impact the security challenges and possible countermeasures in the \ac{iiot}.
These differences particularly concern use and deployment, where the \ac{iiot} requires new security mechanisms beyond the device level, \ie the layered model of the consumer \ac{iot} covers \ac{iiot} security only partially.
In the next section, we therefore summarize the main similarities and differences between consumer and industrial \ac{iot}.

\begin{figure*}
	\centering
	\includegraphics[]{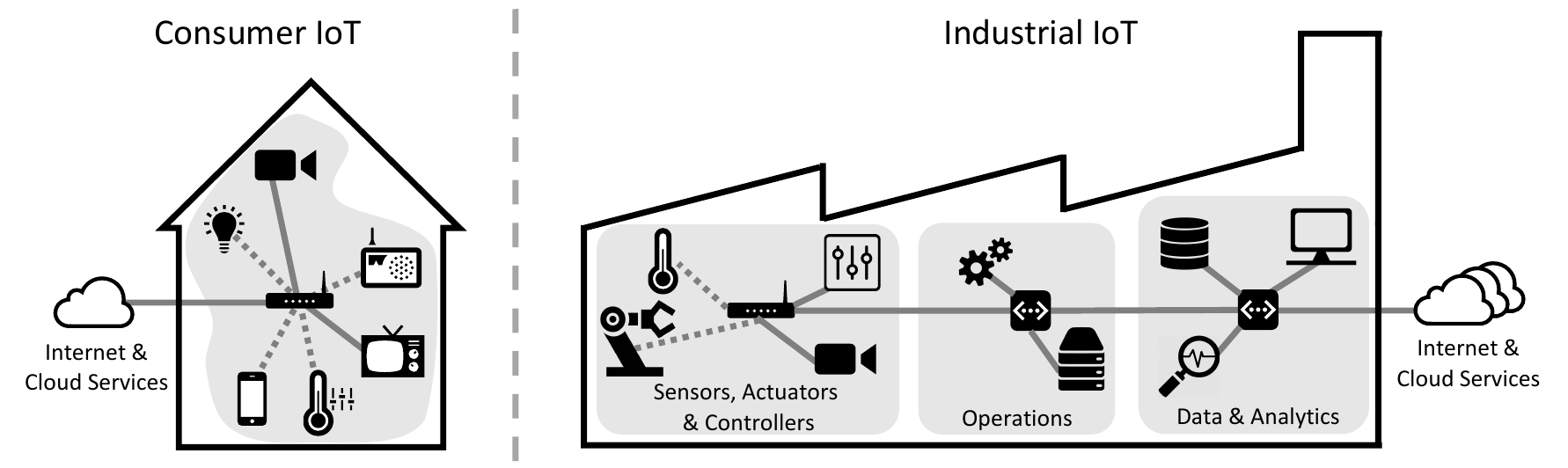}
	\caption{General architecture of the consumer \ac{iot} (left) and the industrial \ac{iot} (right). The consumer \ac{iot} mainly consists of devices that outsource data processing to the cloud. In the \ac{iiot}, in turn, local data processing is complemented with cloud services to optimize processes and to provide new services.}
	\label{fig:iiot-architecture}
\end{figure*}

\section{From Consumer IoT to Industrial IoT}
\label{sec:towards-iiot}

\begin{table*}
  \centering
  \caption{Comparison between consumer and industrial IoT}
  \label{tab:iot-versus-iiot}
  \begin{tabular}{@{}llll@{}}
	\toprule
    \textbf{Category}     &    \textbf{Characteristic}    & \textbf{Consumer IoT}                   & \textbf{Industrial IoT}      \\
    \midrule
    \multirow{2}{*}{Application}  & Service model                        & Human-centered                 & Machine-centered    \\
    & Criticality                          & Not stringent                  & Mission-critical    \\ \hline   
    \multirow{3}{*}{Device} & Number of devices per home/factory & Low to medium                  & Medium to high      \\
    & Lifetime & $3$ to $5$ years & $10$ to $30$ years \\
    & Hardware complexity                    & Low                            & Low to medium       \\ \hline
    \multirow{4}{*}{Data traffic} & Data volume                          & Medium                         & High                \\    
    & Data confidentiality                 & Privacy-oriented               & Business-oriented   \\
    & Traffic type                         & Periodic and event-driven      & Periodic            \\
    & Use of (wireless) communications     & Unstructured, contention-based & Structured, planned \\
    \bottomrule
  \end{tabular}
\end{table*}

There are fundamental differences between the consumer and the industrial \ac{iot}, which have a strong impact on the motivation for attacks and the design and implementation of countermeasures.
Consequently, in this section, we shortly explain the common features and the differences between consumer and industrial \ac{iot}. 
This section thus lays the foundation for the \ac{iiot}-specific security goals and challenges, which we present in the next section, and for providing effective protection mechanisms, which we survey in \pref{sec:iiot-sec}.

\subsection{Common Features of Consumer IoT and Industrial IoT}

Both the private and the industrial sectors benefit from continuous price reductions of software and hardware for computation and communication~\cite{AIMo10, XHLi14} since both domains more and more share the same technologies~\cite{GaHa13}.
This development has led to a large number of pervasive computing devices, which are connected to each other and to external cloud services via the Internet.
This includes, on the one hand, devices that were not connected to the Internet in the past, \eg home appliances in the private domain, or \acp{plc} in the industrial domain.
On the other hand, this also comprises a large number of cheap sensors, which are being deployed to monitor physical processes, \eg to automate heating and cooling in private homes or to support process safety in the industrial domain.
A key enabler for the \ac{iot} is, beyond any doubt, wireless communications increasing flexibility at low deployment and maintenance costs.
Concerning application and deployment, there are, however, considerable differences between industrial and consumer \ac{iot} impacting the security, which we illustrate in the following.

\subsection{Differences Between Consumer IoT and Industrial IoT}
We begin by analyzing the differences between consumer and industrial \ac{iot} by taking a closer look at the respective deployments.
\pref{fig:iiot-architecture} shows a typical consumer \ac{iot} setup (left) compared to a typical \ac{iiot} setup (right).
While devices in the consumer \ac{iot} are, in general, directly connected to the Internet to provide their functions to the users, the industrial \ac{iot} is characterized by a strong interconnection between field devices, controllers, servers, etc. 
That is, a large portion of the data processing happens within the local networks.
The connection to the Internet and cloud services complements local processing by offering enhanced process monitoring and optimizations.
Moreover, with the emerging \ac{iiot}, the typical hierarchical structure of industrial networks, such as described in \cite{CDVa13}, is progressively disbanded due to increased connectivity and gradual merging of \ac{it} and \ac{ot}.

To provide a more detailed comparison between consumer and industrial \ac{iot}, we distinguish between three distinct categories, namely \emph{application}, \emph{device}, and \emph{data traffic}, where each category has different characteristics (\cf \pref{tab:iot-versus-iiot}).

\subsubsection{Application}
The most notable difference concerning the application focus of consumer and industrial \ac{iot} lies in the different service models: 
While the consumer \ac{iot} strives to assist humans in their daily tasks~\cite{AIMo10}, the main objective of the \ac{iiot} is to augment automation tasks by interconnecting \ac{ot} with \ac{it}~\cite{WSJa17}.
The latter is thus dominated by autonomous \ac{m2m} connections for monitoring and control, while the former is mainly triggered by the presence and the routine of human users.
Closely related to these service models is the criticality of the provided services, where the \ac{iiot} also comprises safety- and mission-critical tasks.
In contrast, the consumer domain typically does not impose such stringent requirements.
Ensuring security in the \ac{iiot} is thus an important part of ensuring process safety.

\subsubsection{Device}
Concerning the used devices, the average number of connected devices deployed in a factory is expected to be an order of magnitude higher than in a private home.
Furthermore, in the industrial domain, the lifetime of (specialized and expensive) devices may be up to $30$~years.
In comparison, consumer devices are replaced more regularly after $3$ to $5$~years, on average~\cite{SPL+11, BrSa15}, since such devices are often less complex and cheaper.
Security measures in the \ac{iiot} thus need to be scalable, adaptable, and even retrofittable. 

\subsubsection{Data Traffic}
In both the consumer and the industrial \ac{iot}, the amount of collected data is rapidly increasing, requiring new analytic approaches, sometimes referred to as \emph{Big Data}~\cite{MNG+17}.
While consumer settings reveal rather heterogeneous data traffic depending on the different applications, the \ac{iiot} mainly consists of sensing, monitoring, and control tasks. 
Therefore, in the \ac{iiot}, we observe regular and deterministic data flows, facilitating the implementation of network policies and intrusion detection.

For both domains, there is a need for data confidentiality.
In the industrial domain, this mostly relates to preserving company secrets and customer data, whereas in the consumer \ac{iot}, this relates to the protection of the user's privacy. 

\subsection{Summary}
Both the consumer and the \ac{iiot} share some technological similarities.
Most notably, both domains experience an increase in connectivity enabled by (cheap) wireless communication hardware.
Regarding use and deployment, however, there are considerable differences, as summarized in \pref{tab:iot-versus-iiot}.
Most importantly, the key differences lie in the criticality of the applications, the expected lifetime of the devices, and the determinism as well as periodicity of the data traffic.

Hence, for the \ac{iiot}, we need to rethink security fundamentally to address these findings.
Moreover, most security approaches for the consumer \ac{iot} predominantly focus on securing individual devices and do not explicitly consider their interdependence, deployments, and tasks. 
Therefore, in the next section, we specifically identify the resulting \ac{iiot} security goals and challenges.
 
\section{IIoT Security Goals and Challenges}
\label{sec:iiot-risks}
As a preliminary step towards securing the \ac{iiot}, we first need to outline the specific security goals for \ac{iiot} deployments.
Therefore, we begin with a description of the security requirements (\cf \pref{sec:iiot-risks-requirements}), followed by the considered attacker model (\cf \pref{sec:iiot-risks-attacker}).
Eventually, we summarize the challenges for achieving security in the \ac{iiot} (\cf \pref{sec:iiot-risks-challenges}).

\subsection{Requirements}
\label{sec:iiot-risks-requirements}
In general, the security requirements of the \ac{iiot} are related to those of the consumer \ac{iot}, however, with a different prioritization. 
In particular, security in the industrial domain is closely related to ensuring safety and productivity.
Therefore, the foremost security requirements for \ac{iiot} systems are \emph{availability} and \emph{integrity}~\cite{SWWa15, DNVC05}.

A loss of availability, \eg caused by a \ac{dos} attack, may lead to a production stop or even jeopardize safety measures and consequently endanger human life.
Similarly, missing integrity, \eg manipulated sensor values, may result in corrupted products, waste of resources, or ineffective safety.
In this context, corrupted products may even cause a higher loss of revenue than a (short-term) production stop, because a long time may pass until the sabotage is noticed, let alone the risk of damage claims from clients and contract partners.

Although \emph{confidentiality} is widely considered of secondary importance in the industrial domain, it is becoming increasingly important for the \ac{iiot} to protect the \ac{it}-infrastructure against theft of customer data and industrial espionage, since more and more processes are being digitized.
Moreover, \emph{authenticity}, \emph{authorization}, and \emph{non-repudiation}, which are closely related to each other, further reduce the risk of intrusion and sabotage, while also facilitating post-factum investigations.
Implementing these security goals in industrial environments is, however, challenging due to several reasons, which we further explain in \pref{sec:iiot-risks-challenges}, after specifying the considered attacker model.

\subsection{Attacker Model}
\label{sec:iiot-risks-attacker}
Attacker models are a valuable framework to formally model different security threats and resulting risks, especially in the context of the \ac{iiot}~\cite{adepu2016generalized}.
In contrast to other domains~\cite{ryan2014enhanced}, the specifics of \ac{iiot} deployments demand a realistic and robust attacker model, in which attackers are not restricted from acting entirely malicious.
Consequently, we assume the Dolev-Yao attacker model~\cite{DoYa83}, which is typically used also in the context of (consumer) \ac{iot} security.
The model assumes that an attacker can overhear, intercept, and modify any message in the network, i.e., perform both \emph{passive} and \emph{active} attacks.
The attacker is, however, assumed not to be capable of breaking the applied (state-of-the-art) cryptographic methods.
Besides, for \ac{iiot} security, it is essential to further distinguish between \emph{outside} and \emph{inside} attackers, where the former can only attack \ac{iiot} devices via the network, whereas the latter has additional physical access to the devices.
These considerations enable us to identify challenges for \ac{iiot} security as well as allow us to evaluate to which extent potential countermeasures address identified challenges, risks, and threats.

\subsection{Challenges}
Besides considering the specific \ac{iiot} security requirements and attacker model, there are unique challenges in securing the \ac{iiot}, which we outline in the following.

\label{sec:iiot-risks-challenges}
\textbf{Long-lived components.}
\ac{iiot} devices have a significantly longer lifespan compared to consumer \ac{iot} devices, \cf \pref{tab:iot-versus-iiot}.
This increases the necessity to already consider application and communication security during the development of such devices and, more importantly, to update the software regularly once devices are deployed.
This challenge does, however, not only refer to newly deployed devices; it mainly concerns already deployed devices, which came with little or no security mechanisms and with a cumbersome update mechanism, yet are expected to be operated for decades.
With increased connectivity in the \ac{iiot}, the risk of security breaches also increases, especially when previously isolated legacy components are likewise integrated into the network~\cite{PAEb14}.

\textbf{Large number of devices.}
The \ac{iiot} consists of an increasing number of (mainly resource-constraint) devices.
These devices need to be deployed, configured, and managed in evolving automation tasks.
The sheer number of devices, especially compared to the significantly smaller consumer deployments, thus imposes a need for scalable, automatic approaches not only for operation but also for the implementation and configuration of security measures.   

\textbf{High connectivity.}
The major benefits of the \ac{iiot} are based on strong connectivity between \ac{it} and \ac{ot}, and even to the Internet, enabling more efficient and flexible industrial productions~\cite{LYD+17}.
Given this paradigm, it is increasingly difficult to separate and isolate devices according to their functionality and thus to restrict unauthorized access.
However, according to \acs{nist}~\cite{SPL+11}, network segmentation is a practical approach to protect \acp{ics} and thus needs to be further pursued for the \ac{iiot}.

\textbf{Critical processes.}
A crucial part of \acp{ics} is safety- and mission-critical processes, which do not tolerate outages and thus require high availability.
Moreover, such processes highly depend on data integrity, since even small deviations may pose a safety risk or harm product quality.
Security measures, however, may conflict with these requirements, \eg when they increase the communication and process latency.
Instead of trading security against low latency, adapted security measures for safety- and mission-critical processes are needed.

\textbf{Data confidentiality.}
With the \ac{iiot}, increasing amounts of data are collected, \eg cloud services~\cite{SASa16} use process and meta information for control and optimization.
This data also includes customer data and business secrets, which need to be protected from unauthorized access.
The key challenge is thus to ensure confidentiality while also providing access to authorized \ac{iiot} services for processing and analysis.

\textbf{Human error and sabotage.}
Besides malicious attackers targeting \acp{ics}, there is also the risk of accidental failures, \eg caused by a force majeure or misconfigurations, threatening the system availability. 
This threat is gaining importance in the \ac{iiot}, where the networked components exhibit an increased interdependence.
Similarly, protection against sabotage from the inside, \eg a disgruntled worker~\cite{GaHa13}, remains a challenging task, since access control and isolation are becoming more difficult.
Compared to consumer \ac{iot} deployments, where typically only a small group of people has direct access to \ac{iot} devices, \ac{iiot} deployments may not only be accessed by (a large number of) employees, but also by customers, suppliers, and collaborators~\cite{pennekamp_dataflows_2019}.
Solving these challenges, once again, requires clear access policies and automatic approaches to handle the increasing number of devices and interconnections.

The presented challenges for securing the \ac{iiot} show that new and adapted security measures are needed. 
Existing security solutions for the consumer \ac{iot} thus do not account for the tremendous differences in use and deployment when shifting towards the industrial domain.
In the following, we therefore present security measures for the \ac{iiot} and survey to which extent they address the previously identified challenges.

\section{Securing the IIoT}
\label{sec:iiot-sec}

\begin{table*}
  \centering
  \footnotesize
  \setlength{\tabcolsep}{5pt}
  \caption{Comparison of \ac{iiot} Security Measures}
  \label{tab:discussion}
  \begin{tabularx}{\linewidth}{@{}p{0.01cm}X>{\centering}p{1.76cm}>{\centering}p{1.76cm}>{\centering}p{1.76cm}>{\centering}p{1.76cm}>{\centering}p{1.76cm}>{\centering}p{1.76cm}@{}} 
	\toprule 
    & & \textbf{Long-lived components} & \textbf{Large number of devices} & \textbf{High connectivity} & \textbf{Critical processes} & \textbf{Data confidentiality} & \textbf{Human error \& sabotage} \tabularnewline
    \midrule
    \parbox[t]{2mm}{\multirow{3}{*}{\rotatebox[origin=c]{90}{\scriptsize\S~\ref{sec:cryptography}}}} & \textbf{Tailored cryptography \& authentication} & & & & & & \tabularnewline
    & Lightweight auth.\ and encryption~\cite{HHS+18, EMM+19} & \LEFTcircle & \LEFTcircle & \Circle & \CIRCLE & \CIRCLE & \Circle \tabularnewline
    & Certificateless encryption~\cite{MHK+18, FLC+18} & \LEFTcircle & \CIRCLE & \LEFTcircle & \Circle & \CIRCLE & \Circle \tabularnewline
    \midrule
    \parbox[t]{2mm}{\multirow{3}{*}{\rotatebox[origin=c]{90}{\scriptsize\S~\ref{sec:patch-management}}}} & \textbf{Patch management} & & & & & & \tabularnewline
    & Automated updates~\cite{MMTB19} & \LEFTcircle & \CIRCLE & \Circle & \Circle & \LEFTcircle & \Circle \tabularnewline
    & Detection of vulnerabilities~\cite{FlMu18, GeTh18} & \CIRCLE & \LEFTcircle & \Circle & \LEFTcircle & \LEFTcircle & \LEFTcircle \tabularnewline
    \midrule
    \parbox[t]{2mm}{\multirow{3}{*}{\rotatebox[origin=c]{90}{\scriptsize\S~\ref{sec:service-isolation}}}} & \textbf{Service isolation \& access control} & & & & & & \tabularnewline
    & Trusted execution environments~\cite{LHWi15, PGP+17} & \Circle & \LEFTcircle & \LEFTcircle & \CIRCLE & \CIRCLE & \LEFTcircle \tabularnewline
    & Netw.\ security policies~\cite{LeWe16, HRG+18, HWJa16, RMEl19} & \CIRCLE & \LEFTcircle & \CIRCLE & \Circle & \LEFTcircle & \CIRCLE \tabularnewline
    \midrule
    \parbox[t]{2mm}{\multirow{3}{*}{\rotatebox[origin=c]{90}{\scriptsize\S~\ref{sec:monitoring}}}} & \textbf{Network monitoring \& intrusion detection} & & & & & & \tabularnewline
    & Network-based monitoring~\cite{MUSN12, CBP+15, SSG+16} & \CIRCLE & \LEFTcircle & \Circle & \LEFTcircle & \LEFTcircle & \Circle \tabularnewline
    & Process-aware IDS~\cite{goldenberg_accurate_2013, hadvziosmanovic_plc_2014, ZHX+15, chromik_bro_2018} & \CIRCLE & \LEFTcircle &  \LEFTcircle &  \CIRCLE &  \LEFTcircle & \LEFTcircle \tabularnewline
    \midrule
    \parbox[t]{2mm}{\multirow{3}{*}{\rotatebox[origin=c]{90}{\scriptsize\S~\ref{sec:awareness}}}} & \textbf{Awareness, training \& assessment} & & & & & & \tabularnewline
    & Awareness and training \cite{FrBr14, MaTi16, AGA+17}  & \Circle & \Circle & \Circle & \LEFTcircle & \LEFTcircle & \CIRCLE \tabularnewline
    & Assessment \cite{CSET, BHH+18} & \Circle & \LEFTcircle & \Circle & \LEFTcircle & \LEFTcircle & \LEFTcircle \tabularnewline
    \bottomrule
  \end{tabularx}
\end{table*}

To address these severe security challenges and thus to meet the overarching security goals in the \ac{iiot}, we provide a survey of different security approaches for the \ac{iiot}.
For those approaches that were originally developed exclusively for the consumer \ac{iot}, we explain how they can be adapted or applied in an industrial context.
Finally, we recapitulate the key contributions of each approach and provide a summary of our discussion in \pref{tab:discussion}, where the different security approaches are grouped according to general security measures, which corresponds to the structure of this section.
For each approach, we provide a rating that indicates to what extent it addresses the challenges defined in \pref{sec:iiot-risks}, where \Circle{} represents \emph{no consideration}, \LEFTcircle{} denotes a \emph{partial consideration}, and \CIRCLE{} stands for \emph{full consideration}.

\subsection{Tailored Cryptography and Authentication}
\label{sec:cryptography}

Encryption is a key mechanism to ensure confidentiality of data and, moreover, can be used to achieve authentication and non-repudiation (\cf \pref{sec:iiot-risks}).
However, a significant number of devices in the \ac{iiot} is expected to be resource-constraint, demanding the use of lightweight symmetric-key cryptography approaches instead of computationally more expensive public-key cryptography~\cite{DNVC05}.
Nevertheless, symmetric-key cryptography typically lacks a secure and scalable key management infrastructure, and it is challenging to maintain the secrecy of join keys for a large number of participants~\cite{CNL+11}.
Moreover, for safety- and mission-critical processes, both public-key and symmetric-key cryptography typically yield unacceptable delays, which often prevents factory operators from implementing encryption and authentication at all.
From a different perspective, the increased data exchange between devices in the \ac{iiot} and the growing dependence on cloud services demands adequate data protection against unauthorized access.
Therefore, new approaches to cryptography and authentication are required that are specifically tailored to the \ac{iiot}.

To reduce the latency and thus to enable \emph{lightweight authentication and encryption} in industrial communication scenarios, different approaches considering resource-constraint devices have been proposed~\cite{HHS+18, EMM+19}.
For~\cite{HHS+18}, the underlying assumption is that communication in the \ac{iiot} consists mostly of periodic connections with static communication partners.
Consequently, it is possible to partly precompute symmetric encryption and authentication.
More specifically, the authors show that antedated encryption and data authentication with templates significantly reduces security processing by up to \SI{76}{\%}, depending on the packet size.
In turn, \cite{EMM+19} relies on a lightweight authentication scheme, based only on hash and XOR operations, to enable authentication of resource-constraint devices. 
This approach requires, however, that such devices are equipped with a secure element, which might not be the case for legacy devices.

Other approaches specifically focus on securing the communication to external entities such as cloud services~\cite{MHK+18, FLC+18}, where \cite{MHK+18} builds on \cite{YJCZ14} by reducing the computational costs about \SI{30}{\%}.
These approaches are based on \emph{certificateless searchable public-key encryption}, providing simple key management for a large number of devices.
The basic idea is that data is encrypted prior to offloading it to a cloud service and that the encrypted data is searchable such that individual data items are only decrypted \emph{after} they have been retrieved from the cloud.
Such schemes are thus particularly useful when relying on external cloud providers where the confidentiality of outsourced data cannot be guaranteed while addressing an increasing number of devices and connections.

\subsection{Patch Management}
\label{sec:patch-management}

For many (consumer) \ac{iot} devices, manufacturers do either not provide security patches or the installation of a patch requires great manual effort, which prevents users from performing updates.
This situation has led to a large amount of deployed \ac{iot} devices with known vulnerabilities~\cite{PGC+14}.
For a secure \ac{iiot}, patching all used devices in time is indispensable to close vulnerabilities and thus to reduce the risk of attacks~\cite{BCGN16}.
In many companies, however, the internal processes of patching vulnerabilities swiftly, without having to wait for the next planned maintenance window, must be improved~\cite{HSHO18}.
This requires that the respective manufacturers regularly offer security patches for all their devices for an extended lifetime that such devices typically experience in industrial deployments (compared to consumer deployments).

Moreover, automatic approaches for installing patches can facilitate this process for a large number of expected \ac{iiot} devices.
Nevertheless, patching industrial systems typically entails an extensive testing phase prior to the installation to make sure that the patch is compatible with the current setup.
Indeed, NIST recommends performing regression testing as part of a systematic patch management process~\cite{SPL+11}, to ensure safety and to reduce the risk of process downtimes.

In the \ac{iot} context, the IETF working group on Software Updates for Internet of Things~(SUIT) proposes an \emph{automated firmware update mechanism} for resource-constraint devices~\cite{MMTB19}.
This mechanism ensures for each update a standardized description of the involved entities, security threats, and assumptions, as well as a protected end-to-end transmission of the new firmware to the respective devices.
The goal is thus to simplify updating of \ac{iot} devices while, at the same time, providing a secure and standardized procedure. 
For \ac{iiot} devices, this standard can be easily adopted, and, even further, industrial operators can decide only to deploy devices implementing the aforementioned standard. 

Additionally, there are approaches to \emph{detect security bugs and vulnerabilities} actively in \ac{iiot} deployments~\cite{FlMu18, GeTh18}, either by testing \ac{iiot} devices during their idle times or by analyzing vulnerabilities based on an \ac{iiot} network graph.
The performance evaluation of \cite{FlMu18} shows that the use of idle times does not impact the industrial processes, which makes it particularly useful for safety- and mission-critical processes.
Such approaches are a first step towards identifying existing security flaws and their implications on other systems, as well as developing viable countermeasures, \eg by isolating vulnerable devices.

\subsection{Service Isolation and Access Control}
\label{sec:service-isolation}

Even when regularly patching \ac{iiot} devices, the existence of vulnerabilities cannot be completely ruled out, since manufactures might not yet be aware of certain security flaws in their products, which are commonly referred to as \emph{zero-day} vulnerabilities. 
Moreover, the support of deployed legacy devices might be discontinued by their manufactures, and thus security patches are no longer provided. 
In these cases, additional protection mechanisms are needed to prevent attacks against these devices and subsequent attacks to other connected devices.
This includes a defense-in-depth architecture, as proposed by \acs{nist}~\cite{SPL+11}, minimizing the impact of an attack with (internal) firewalls and demilitarized zones.
Additionally, more fine-grained security policies restricting access to computing and network resources for each device and even within a device for single applications and tasks, according to the respective needs, can further reduce the risk of attacks.

Recently, different approaches~\cite{LHWi15, PGP+17} specifically address the application of hardware-security technologies, such as \emph{trusted execution environments}, in industrial use cases.
The major challenge for considering technologies like ARM TrustZone or security controllers in the context of the \ac{iiot} is to support also time-critical applications.
First prototypical evaluations show that even resource-constraint devices can securely execute safety- and mission-critical tasks with the help of trusted execution environments.
ARM TrustZone, for example, outperforms security controllers regarding the processing time by, at least, an order of magnitude due to a more efficient data handling according to~\cite{LHWi15}.
Such approaches are, nonetheless, only applicable for future devices generations where the respective hardware is available.

A viable solution for legacy and non-patchable systems is to enforce \emph{security policies within the network}. 
Besides offering additional protection against unauthorized access to such devices, this is also an effective way to prevent follow-up attacks from already infected devices to other parts of the network.
The IETF has recently proposed \ac{mud}~\cite{LeWe16}, where the \ac{iot} device manufacturer defines networking rules based on the functionality of the respective device, \ie most \ac{iot} devices have a very specific purpose and thus do not need full network access to fulfill their tasks.
A central enforcer within the local network then blocks all connections that do not comply with the defined \ac{mud} rules and thus limits the opportunities for attacks.
Current research in this area~\cite{HRG+18} has shown that automatic approaches can be leveraged to derive \ac{mud} rules and that this approach hence supports already deployed devices, even when the manufacturers themselves do not provide the respective rules.
This approach can be further realized with a \ac{sdn} approach~\cite{HWJa16, RMEl19} facilitating the implementation of such policies in evolving industrial networks.
Especially in \ac{iiot} networks, the property of mostly periodic connections with static communication partners~\cite{HHS+18} can be leveraged to restrict communication capabilities further.

\subsection{Network Monitoring and Intrusion Detection}
\label{sec:monitoring}

Intrusion detection mechanisms are indispensable to detect malicious activities, \eg ongoing attacks, and thus to prevent further damage.
They are especially relevant when the preventive security measures have been implemented inadequately or when attackers exploit zero-day vulnerabilities.
Existing \acp{ids} for traditional \ac{it} networks can, unfortunately, not simply be applied in the industrial domain~\cite{ZHX+15}.
\acp{ics}, for instance, are dominated by real-time processes and resource-constraint devices, which are less common in traditional \ac{it} networks. 
Likewise, industrial networks often require more vantage points for \ac{ids} monitors, as not all traffic flows through one central point. 
Apart from these challenges, there are also some opportunities when implementing \acp{ids} in the \ac{iiot}:
From the deterministic industrial tasks result rather regular network traffic patterns facilitating anomaly detection compared to the random communication in \ac{it} networks~\cite{HHS+18}.

A prerequisite for intrusion detection in the \ac{iiot} is efficient \emph{network-based monitoring} of the communication traffic.
Research in this direction~\cite{MUSN12, CBP+15} focuses on passive, real-time capable approaches that do not interfere with critical industrial processes.
Moreover, the costs for inspecting packets must be reduced to ensure scalability with an increasing number of devices.
In the context of the consumer \ac{iot}, a promising approach is to monitor traffic at the flow-level instead of inspecting every packet~\cite{SSG+16}.
This approach thus presents a viable solution to address scalability in the \ac{iiot}, since the evaluation shows that the amount of monitored traffic could be reduced by an order of magnitude.

A further step towards anomaly and intrusion detection in the \ac{iiot} is to consider, besides the network behavior, also the industrial process, which is referred to as \emph{process-aware \ac{ids}}~\cite{goldenberg_accurate_2013, hadvziosmanovic_plc_2014, ZHX+15, chromik_bro_2018}.
This requires a model of the respective industrial processes, \eg by using domain expertise or machine-learning techniques.
The advantage of such holistic approaches is that detection error rates can be reduced further, \eg the evaluation results of \cite{ZHX+15} show an achieved accuracy of \SI{99.82}{\%} where, additionally, attacks are distinguished from system faults.
Hence, to some extent, such approaches may also be used to detect configuration errors made by operators.

\subsection{Awareness, Training, and Assessment}
\label{sec:awareness}

The growing prevalence of the \ac{iiot} emphasizes that \ac{it} security affects all areas of a company and therefore requires deep roots in the corporate culture~\cite{Wood07}.
Successful protection against attacks hence does not exclusively rely on technology; it significantly depends on humans, \eg workers, managers, etc., and on the implementation of security policies and procedures within the company.
Especially the larger group of people interacting with \ac{iiot} deployments compared to consumer \ac{iot} settings renders awareness, training, and assessment challenging.

A preliminary step to embed security in the corporate culture is to create \emph{cyber situational awareness}~\cite{FrBr14}, \ie being aware of the potential security threats and risks and recognizing the need for implementing security measures.
This includes the business case for technical tools (\eg \acp{ids} and access logs), appropriate resources (\eg incident response teams), and \emph{security training} of employees.
The latter educates employees about complying with security policies and how to react in security-critical situations, \eg an attack.
For a first informative approach, the OWASP Foundation provides, in the context of the (consumer) \ac{iot},  general security guidelines for manufacturers, developers, and users~\cite{OWASP17}, which also carry over to \ac{iiot} devices.
To understand the impact of attacks and the protection against security threats in an industrial context, several testbeds for security in \acp{ics} have emerged, \eg \cite{MaTi16, AGA+17}.
The advantage of such experimental training approaches is the practical experience that the participants gain through direct interaction.
Furthermore, \cite{BaMa19} shows that a simple transfer of knowledge is not enough to change user behavior. 
Instead, security training needs to be hands-on and provide continuous feedback for participants to ensure long-term changes in their behavior. 

Finally, a regular \emph{security assessment} of the current deployment is indispensable for determining if the implemented security measures are adequate and correct.
There are various tools, \eg the Cyber Security Evaluation Tool~\cite{CSET} or the \ac{iiot} analysis framework~\cite{BHH+18}, which facilitate the security assessment of larger deployments and are thus particularly relevant to strengthen and evaluate security in the \ac{iiot}.

\subsection{Summary and Recommendations}
\label{sec:discussion}

In summary, the security challenges we identified in \pref{sec:iiot-risks-challenges} may be addressed as follows (\cf \pref{tab:discussion}):
The challenge of \textbf{long-lived components} requires automatic patch management and vulnerability detection (\cf \pref{sec:patch-management}), which should be complemented by access control (\cf \pref{sec:service-isolation}) and network monitoring (\cf \pref{sec:monitoring}) since legacy devices might not receive security patches.
Additionally, since the \ac{iiot} consists of a \textbf{large number of devices}, security needs to be scalable, \eg using certificateless encryption (\cf \pref{sec:cryptography}).
The \textbf{high connectivity} of the \ac{iiot} components could be addressed with individual access control policies, restricting devices according to the \emph{least privilege} approach (\cf \pref{sec:service-isolation}).
For \textbf{critical processes}, it is of paramount importance to use lightweight authentication and encryption adhering to the latency requirements (\cf \pref{sec:cryptography}), as well as latency-aware trusted execution environments (\cf \pref{sec:service-isolation}).
The use of encryption is indispensable to achieve \textbf{data confidentiality}, especially when data is stored on external cloud servers (\cf \pref{sec:cryptography}).
In this context, service isolation can also contribute to a higher protection of confidentiality (\cf \pref{sec:service-isolation}).
Finally, \textbf{human error and sabotage} must be prevented with clear access policies and regular awareness training (\cf \pref{sec:awareness}).
Such policies and training include the protection against inside attackers (\cf \pref{sec:iiot-risks-attacker}) as well as outside attackers, \eg using \emph{social engineering} to gain unauthorized access.

\section{Related Work}
\label{sec:related-work}

\begin{table}
  \centering
  \caption{Summary of recent IIoT surveys}
  \label{tab:related-work}
  \begin{tabular}{@{}p{0.35\columnwidth}p{0.55\columnwidth}@{}}
    \toprule
    \textbf{Category} & \textbf{Summary and Impact} \\
    \midrule
    Literature studies focusing on the specific challenges and requirements of the \ac{iiot} \cite{SWWa15,TDFD19,lezzi_cybersecurity_2018,yu_survey_2019}. & This work partially builds on the identified challenges and requirements (\cf \pref{sec:iiot-risks}). Subsequently, this work also discusses possible countermeasures (\cf \pref{sec:discussion}). \\
    \midrule
    Increased connectivity in the \ac{iiot} \cite{SASa16, pennekamp_dataflows_2019}. & Confdentiality in the \ac{iiot} also concerns local networks and protection against inside attackers (\cf \pref{sec:service-isolation} \& \pref{sec:monitoring}). \\
    \midrule
    The security impact of industrial processes \cite{KBLa18, CRFF17, PWS+17}. & The identified weaknesses must be addressed with appropriate countermeasures taking into account long-lived components (\cf \pref{sec:iiot-risks-challenges} \& \pref{sec:discussion}). \\
    \bottomrule
  \end{tabular}
\end{table}

We are not first to survey security challenges in the \ac{iiot}.
Different surveys were published covering different aspects of \ac{iiot} security, which can be roughly classified into \emph{literature studies on challenges and requirements}, \emph{increased connectivity in the \ac{iiot}}, and \emph{the security impact of industrial processes}.
In the following, we shortly summarize recent related work according to these categories (\cf \pref{tab:related-work}) and explain how our work complements and extends the current state-of-the-art.

\subsection{(Systematic) Literature Studies on IIoT Security}
Several literature studies focus on the specific security challenges and requirements of the \ac{iiot}, considering the characteristics and dependencies of \ac{iiot} components, \eg \cite{SWWa15,TDFD19,lezzi_cybersecurity_2018,yu_survey_2019}.
Especially those surveys that primarily rely on systematic or structured literature studies~\cite{TDFD19,lezzi_cybersecurity_2018} are well-suited to gain a comprehensive picture of the research landscape concerning \ac{iiot} security.
While \cite{yu_survey_2019, TDFD19} mainly elaborate on the \ac{iiot} security challenges and requirements (similar to \pref{sec:iiot-risks}), \cite{SWWa15, lezzi_cybersecurity_2018} also mention possible countermeasures. 
However, these surveys leave open to what extent the presented countermeasures meet the requirements and solve the respective security challenges (\cf \pref{sec:discussion}).
In our work, we partially rely on the identified challenges and requirements (\cf \pref{sec:iiot-risks}).
We then, however, go one step further and specifically identify possible countermeasures and discuss how they address the different requirements and security challenges (\cf \pref{sec:discussion}).

\subsection{Increased Connectivity in the IIoT}
Recent research addresses the security issues that arise from the increased connectivity in the \ac{iiot}, \ie more and more \acp{ics} relying on cloud services and the sharing of data between different organizations, \eg \cite{SASa16, pennekamp_dataflows_2019}, emphasizing the need for confidentiality (\cf \pref{sec:iiot-risks-challenges}).
In particular, \cite{SASa16} presents security challenges and countermeasures of cloud-based \acp{ics}.
The authors of~\cite{pennekamp_dataflows_2019}, in turn, describe the challenges when different stakeholders exchange production data, \ie customers, suppliers, and manufacturers.
These approaches, however, do not specifically focus on the important challenge that a large part of the increased connectivity in the \ac{iiot} concerns the local networks of a manufacturer and protection against inside attackers.
Thus, they lack a discussion about security measures at the network layer (\cf \pref{sec:service-isolation} \& \pref{sec:monitoring}).
We fill this gap by also considering local networks in the \ac{iiot} and corresponding countermeasures to protect against inside attackers.

\subsection{Impact of Industrial Processes on Security}
Finally, a prominent stream of \ac{iiot} security research considers the impact of the respective industrial process on security, \eg \cite{KBLa18, CRFF17, PWS+17}.
More specifically, \cite{KBLa18} identifies the strong coupling between safety and security in the \ac{iiot}, \ie ensuring safety requires security, where security, in some cases, conflicts with safety.
Hence, the authors of~\cite{CRFF17} propose to identify the security challenges of the \ac{iiot} according to the typical product lifecycle, \ie from design to maintenance.
Their work thus underlines the importance of considering the respective challenges at each stage and the respective countermeasures. 

Similarly, the authors of~\cite{PWS+17} present a taxonomy for attacks in the \ac{iiot}, where they especially highlight the connection of vulnerabilities to the impact of the attacks on the industrial processes.
These approaches thus focus on providing a practical methodology to better identify vulnerabilities in a real \ac{iiot} deployment.
We complement these efforts by providing a discussion about possible countermeasures and their feasibility to address these weaknesses, especially when considering long-lived components and hence retrofittable approaches (\cf \pref{sec:iiot-risks-challenges} \& \pref{sec:discussion}). 

In summary, we complement the efforts of other surveys in the area of \ac{iiot} security by providing a mapping from identified security challenges to corresponding countermeasures.
Here, we specifically focus on the unique challenges of \ac{iiot} deployments, especially those resulting from long-lived components and increased connectivity.
Our work thus combines the efforts of previously independently considered approaches into a comprehensive survey of challenges \emph{and} opportunities in securing the \ac{iiot}.

\section{Conclusion}
\label{sec:conclusion}

In this paper, we analyze requirements and approaches for securing the currently emerging \ac{iiot} as part of the move towards Industry~4.0.
Based on the well-researched security issues of the consumer \ac{iot}, we identify fundamental differences compared to the industrial domain, which lead to unique security goals and challenges for securing the \ac{iiot}.
To address these, we review recent research efforts and best practices concerning their applicability and effectiveness, considering the previously identified challenges.
Our analysis shows that a wide range of countermeasures is needed and that those that were initially tailored to consumer settings need to be adapted to also work in industrial settings.
Albeit the substantial challenges that need to be considered for securing the \ac{iiot}, such as coping with long-lived components and high connectivity, there are also unique opportunities that can be leveraged for the effective implementation of security measures.
The regular and predictable communication flows, for example, allow to specify clear network policies and facilitate the implementation of automatic intrusion detection.
Based on our analysis, we identify the following lessons learned for future work:
\begin{itemize}
\item[(i)] Security mechanisms for the \ac{iiot} need to be tailored to resource-constraint devices as well as (time) critical industrial processes. 
Future work must hence further explore security protocols with a negligible impact on the process latency.
\item[(ii)] Secure configurations of long-lived components and legacy devices that cannot be patched require complementary security mechanisms at the network layer, such as access policies and monitoring.
In this context, a promising research direction is network traffic policies for \ac{iiot} devices and process-aware \acp{ids}.
\item[(iii)] The risk of inside attackers should not be underestimated and should be additionally addressed by awareness and regular training of employees. 
Research should, therefore, particularly focus on developing and improving \ac{iiot} security training methods, \eg by gamification.
\end{itemize}

However, \ac{iiot} and Industry~4.0 are still in their infancy, where the promised disruptive changes are yet to come in the near future. 
Therefore, future research must address, beyond improving traditional security measures, new approaches to overcome the challenges of the emerging \ac{iiot}.
In this regard, one important stream of research are \ac{dlt} and blockchains~\cite{WLI+19}, offering immutable and decentralized accountability in heterogeneous scenarios.
Additionally, smart contracts~\cite{ZKS+19} are a promising application of blockchains to distributively enforce access control in industrial environments.
The balancing act of \ac{iiot} security hence lies in leveraging such promising technology for new device generations with novel use-cases, as well as considering existing long-lived deployments with legacy hardware.

\begin{IEEEbiography}[{\includegraphics[width=1in,height=1.25in,clip,keepaspectratio]{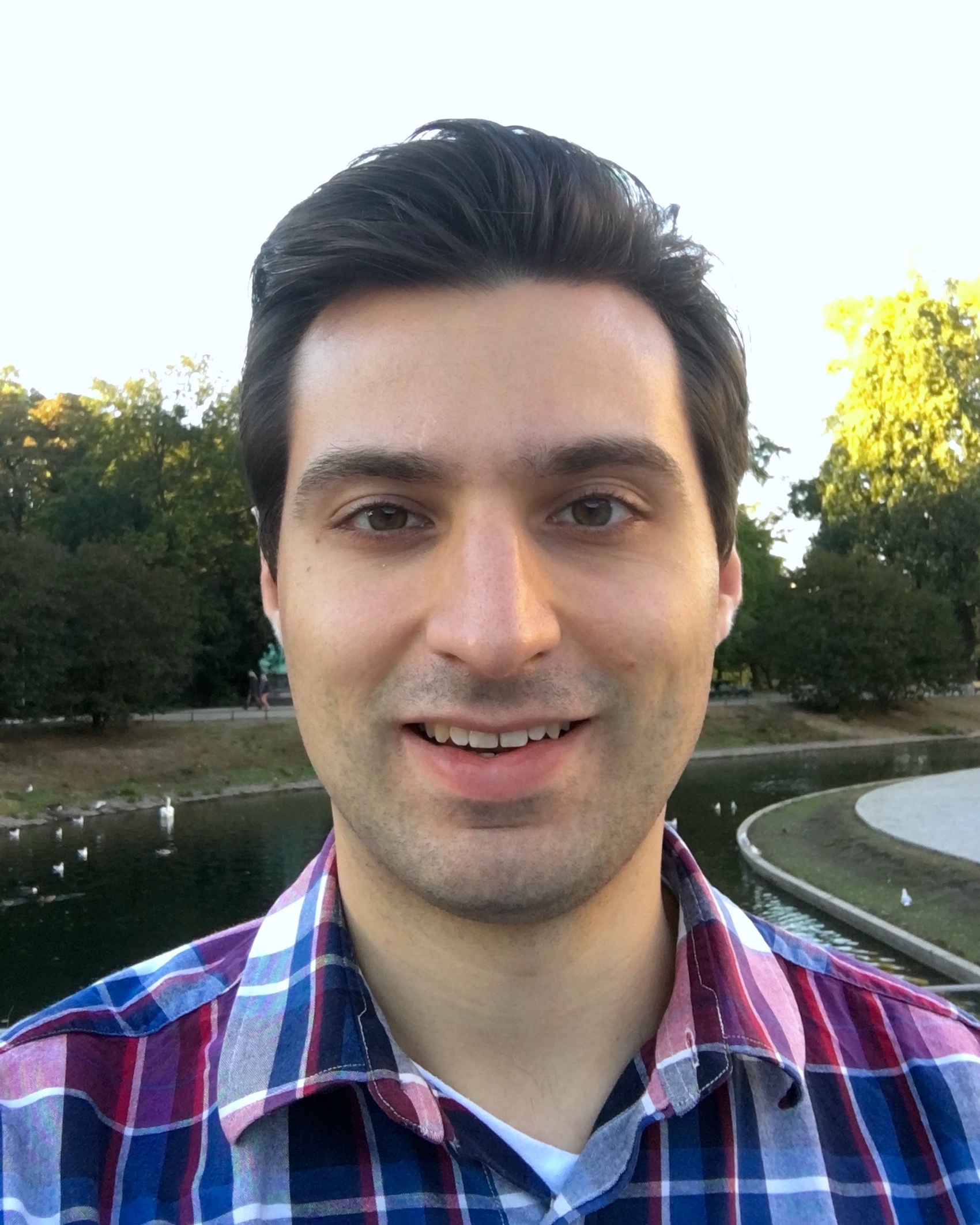}}]{Martin Serror}
received his B.Sc.\ degree in Computer Science at RWTH Aachen University in 2012.
Afterward, he studied abroad at Universitat Politècnica de València, Spain, for two semesters.
He returned to RWTH Aachen University and obtained his M.Sc.\ degree in computer science in 2014 (with honors).
He is currently working toward a Ph.D.\ degree at the Chair of Communication and Distributed Systems (COMSYS).
His research interests lie mainly in dependable wireless systems for critical M2M communications, and IIoT security.
\end{IEEEbiography}

\begin{IEEEbiography}[{\includegraphics[width=1in,height=1.25in,clip,keepaspectratio]{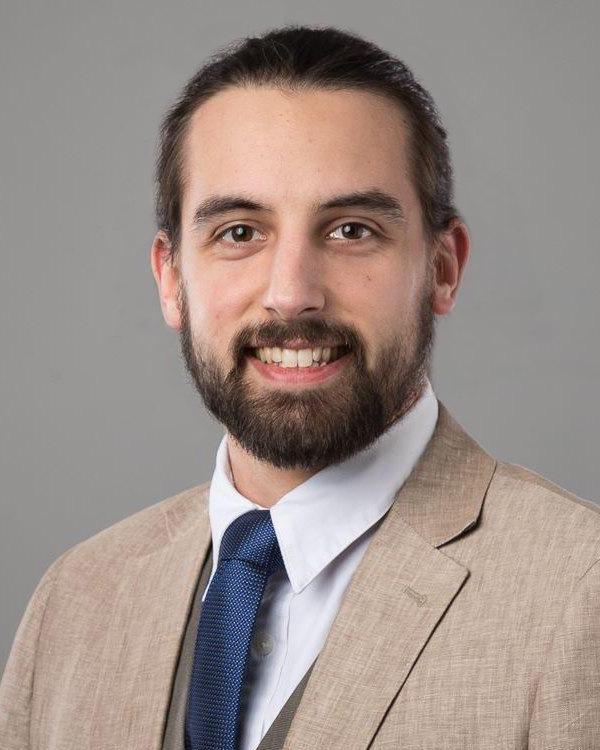}}]{Sacha Hack}
is a PhD student and research assistant in the field of data networks, IT security and digital forensics at the University of Applied Sciences Aachen. 
In 2018 he graduated with a master’s degree in computer science (M.Eng.\ Infomation Systems Engineering). 
He is currently working in the CONSENT project and is concerned with IT security of IoT devices in the smart home area. 
Since 2019 he has been working part-time as a consultant and training manager in the field of IT security for various companies.
\end{IEEEbiography}

\begin{IEEEbiography}[{\includegraphics[width=1in,height=1.25in,clip,keepaspectratio]{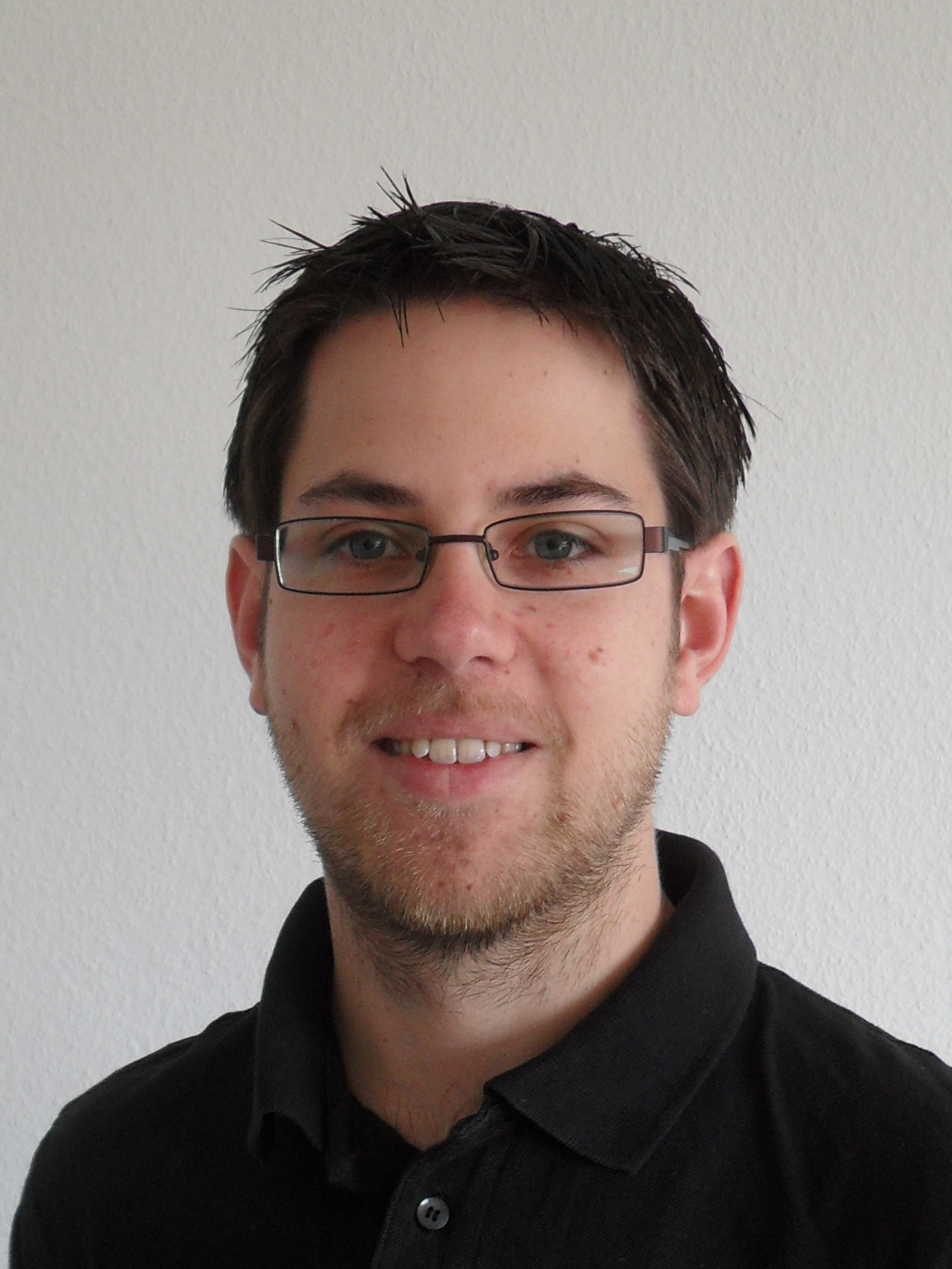}}]{Martin Henze}
received the Diploma (equiv.\ M.Sc.) and PhD degrees in Computer Science from RWTH Aachen University.
He is a postdoctoral researcher at the Fraunhofer Institute for Communication, Information Processing and Ergonomics FKIE, Germany.
His research interests lie primarily in the area of security and privacy in large-scale communication systems, recently especially focusing on security challenges in the industrial and energy sectors.
\end{IEEEbiography}

\begin{IEEEbiography}[{\includegraphics[width=1in,height=1.25in,clip,keepaspectratio]{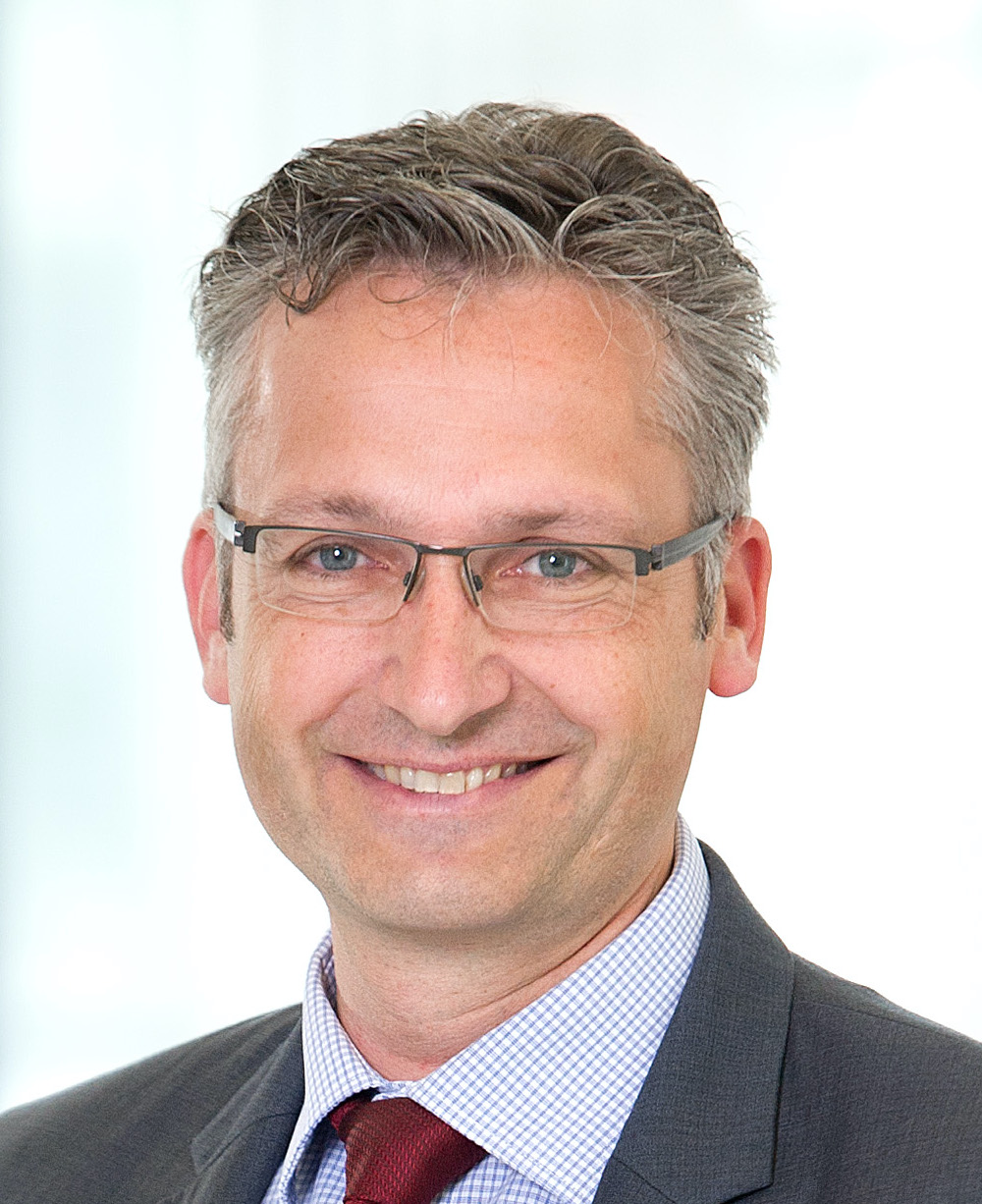}}]{Marko Schuba}
received his Ph.D.\ in Computer Science from RWTH Aachen University. 
He has more than 20 years of experience in IT security-related fields. 
Currently, he is working as a full professor at Aachen University of Applied science, Germany, teaching and doing research in IT security and digital forensics with a focus on industrial control systems, automotive and smart buildings. 
Marko collaborates with a number of companies and authorities, including the State Office of Criminal Investigation NRW, the Federal Criminal Police Office, and Interpol. 
He is also a co-founder of two IT security companies, transferring applied research results into practice.
\end{IEEEbiography}

\begin{IEEEbiography}[{\includegraphics[width=1in,height=1.25in,clip,keepaspectratio]{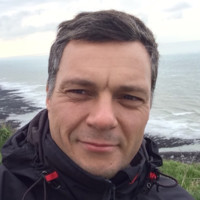}}]{Klaus Wehrle}
received the Diploma (equiv.\ M.Sc.) and Ph.D.\ degrees from University of Karlsruhe (now KIT), both with honors.
He is full professor of Computer Science and Head of the Chair of Communication and Distributed Systems (COMSYS) at RWTH Aachen University.
His research interests include (but are not limited to) engineering of networking protocols, (formal) methods for protocol engineering and network analysis, reliable communication software, as well as all operating system issues of networking. 
He was awarded Junior Researcher of the Year (2007) by the German Association of Professors (Deutscher Hochschulverband).
He is member of IEEE, ACM, SIGCOMM, GI (German Informatics Society), VDE, GI/ITG-Fachgruppe KuVS, the main scholarship board of Humboldt Foundation, the Scientific Directorate of Dagstuhl, and the German National Academy of Science and Engineering (ACATECH).
\end{IEEEbiography}

\end{document}